\documentclass[]{aa}
\input{epsf}
\usepackage{amssymb}

\newbox\grsign \setbox\grsign=\hbox{$>$} \newdimen\grdimen \grdimen=\ht\grsign
\newbox\simlessbox \newbox\simgreatbox
\setbox\simgreatbox=\hbox{\raise.5ex\hbox{$>$}\llap
     {\lower.5ex\hbox{$\sim$}}}\ht1=\grdimen\dp1=0pt
\setbox\simlessbox=\hbox{\raise.5ex\hbox{$<$}\llap
     {\lower.5ex\hbox{$\sim$}}}\ht2=\grdimen\dp2=0pt

\def\mathbi#1{\textbf{\em #1}}

\begin{document}

\title{Interloper treatment in dynamical modelling of galaxy clusters}
\author{R. Wojtak\inst{1} \and E. L. {\L}okas\inst{1} \and
G. A. Mamon\inst{2,3} \and S. Gottl\"ober\inst{4} \and F. Prada\inst{5} \and
M. Moles\inst{5}}
\institute{
Nicolaus Copernicus Astronomical Center, Bartycka 18, 00-716 Warsaw, Poland \and
Institut d'Astrophysique de Paris (UMR 7095: CNRS and Universit\'e Pierre \& Marie Curie),
98 bis Bd Arago, F-75014 Paris, France \and
GEPI (UMR 8111: CNRS and Universit\'e Denis Diderot), Observatoire de Paris, F-92195 Meudon, France \and
Astrophysikalisches Institut Potsdam, An der Sternwarte 16, 14482 Potsdam, Germany \and
Instituto de Astrof{\'\i}sica de Andalucia (CSIC), Apartado Correos 3005, E-18080 Granada, Spain
}

\abstract{}
{
The aim of this paper is to study the efficiency of different approaches to interloper treatment in dynamical
modelling of galaxy clusters.
}
{Using cosmological$N$-body simulation of standard $\Lambda$CDM model, 
we select 10 massive dark matter haloes and use their particles to emulate mock
kinematic data in terms of projected galaxy positions and velocities as
they would be measured by a distant observer. Taking advantage of the full 3D information 
available from the simulation, we select samples of interlopers defined with different criteria. 
The interlopers thus selected provide means to assess the efficiency of different interloper
removal schemes found in the literature.
}
{
We study direct methods of interloper removal based on dynamical or statistical
restrictions imposed on ranges of positions and velocities available to cluster
members. In determining these ranges, we use either the velocity dispersion
criterion or a maximum velocity profile. We also generalize the common approaches taking
into account both the position and velocity information. Another criterion is based on the dependence of
the commonly used virial mass and projected mass estimators on the presence of
interlopers. We find that the direct methods exclude on average 60-70 percent of unbound
particles producing a sample with contamination as low as 2-4 percent.
Next, we consider indirect methods of interloper treatment which are
applied to the data stacked from many objects. In these approaches, interlopers are
treated in a statistical way as a uniform background which modifies the
distribution of cluster members. Using a Bayesian approach, we reproduce the properties of
composite clusters and estimate the probability of finding an interloper as a function of
distance from the object centre.
}
{}


\maketitle

\titlerunning{Interloper treatment}
\authorrunning{R. Wojtak}

\section{Introduction}

The modelling of galaxy kinematics in clusters remains one of the major tools in determining their
properties, in particular their mass distribution and dark matter content. Due to projection
effects, any cluster kinematic data sample inevitably contains galaxies that are not bound to the
cluster and therefore are not good tracers of its gravitational potential. We will call these
galaxies {\em interlopers}. An essential step in
dynamical modelling of clusters by any method is therefore to remove such interlopers from the samples
or take their presence into account statistically. Velocity information can be used to remove obvious
interlopers that are thousands of km s$^{-1}$ off the mean cluster velocity, but there remain numerous
interlopers that lie in a similar general velocity range as the cluster members. Some hints
can be provided by studying the
photometric properties of galaxies or restricting the samples to elliptical galaxies but these
approaches usually do not solve the problem completely.

It has long been recognized that the line-of-sight velocity distribution of galaxies in clusters is
close to a Gaussian. The first attempts to design a scheme to remove the interlopers were based on
this property. Yahil \& Vidal (1977) proposed to calculate the line-of-sight
velocity dispersion of the galaxy
sample, $\sigma_{\rm los}$, and iteratively remove outliers with velocities larger than
$3 \sigma_{\rm los}$. This
simple approach is still widely used today. With enough galaxies in a sample, one can take into account
the dependence of $\sigma_{\rm los}$ on the projected distance from the cluster centre $R$
and perform the rejection
procedure in bins with different $\sigma_{\rm los}$ or fit a simple solution of the
Jeans equation to the
measured line-of-sight velocity dispersion profile, $\sigma_{\rm los} (R)$,
and reject galaxies outside the $3 \sigma_{\rm los} (R)$ curves ({\L}okas et al. 2006).

Perea et al. (1990) discussed another method relying on iterative removal of
galaxies whose absence in the sample causes the biggest change in the mass estimator.
Zabludoff et al. (1990), Katgert et al. (1996) and Fadda et al. (1996) advertised
the use of gaps in the velocity distribution as a way to separate interlopers from
real cluster members. Diaferio \& Geller (1997) and Diaferio (1999) proposed the use of
caustics where the projected distribution function is sufficiently low to separate cluster members
from the surrounding medium. Prada et al. (2003) discussed the solution to the problem
based on the use of escape velocities. The first methods that combine the information
on the position and velocity of a galaxy were proposed by den Hartog \& Katgert (1996)
and Fadda et al. (1996). All these methods aim at cleaning the galaxy sample from
non-members before attempting the proper dynamical analysis of the cluster;
we call them {\em direct\/} methods of interloper removal.

A completely new approach to interloper treatment was pioneered by van der Marel et al. (2000) where,
for the first time, the interlopers were not identified and removed from the sample, but their presence
was taken into account statistically by appropriate modification of the distribution function of the
galaxies. A similar approach was also considered by Mahdavi \& Geller (2004) with more realistic
assumptions concerning the distribution of interlopers. Prada et al. (2003) studied the distribution
of satellites around giant galaxies by fitting to the projected velocity
distribution the sum of a Gaussian and a uniform distribution taking care of the
background. We will refer to this type of methods as {\em indirect\/}. It should be noted that
these methods are mainly applicable to composite clusters, i.e. data sets created by combining
kinematic data from many objects because only then the samples are numerous enough to provide
useful constraints on the interloper fraction.

The different methods of interloper treatment found in the literature are difficult to compare. Each
one of them has a different set of underlying assumptions. They also differ by the amount of parameters
that have to be put in by hand. Most of the methods are iterative and some may not converge. The
ultimate comparison between the methods can only be performed by resorting to $N$-body simulations
where full 3D information is available and true interlopers can be identified. Such tests have
been already attempted (e.g. by Perea et al. 1990; den Hartog \& Katgert 1996; Diaferio 1999).
In particular, van Haarlem et al. (1997) compared the methods of den Hartog \& Katgert (1996)
and Yahil \& Vidal (1977) in terms of the quality of reproduction of the real velocity dispersion.
However, more systematic
study of different procedures is still needed and this is the aim of the present paper. We implement
and generalize different prescriptions for interloper removal found in the literature and apply them
to mock kinematic data created from the simulation. Our goal is to measure the efficiency of the
different methods by measuring fractions of interlopers they remove.

Our choice of methods will of course
be arbitrary. We tried to focus on those easiest to implement, most widely used in the literature
and with the smallest number of pre-selected parameters so that they are applicable not only to
galaxy clusters but to all astronomical systems where kinematic measurements of discrete tracer
can be made (e.g. dwarf spheroidal galaxies). In the near future we plan to apply the methods
discussed here to nearby clusters from the WINGS survey (Fasano et al. 2006) where about 300 redshifts
per cluster will be available.

The problem of the treatment of interlopers is directly related to the problem of the mass estimation
in gravitationally bound objects. We will demonstrate in section 2 that using contaminated
kinematic samples can lead to serious errors in the estimated mass. In addition, several of the
interloper removal schemes we discuss make use of some crude mass estimators. However, the purpose
of this work is not to provide the best method for mass estimation in galaxy clusters. Instead,
we focus on a much narrower issue of how to obtain a clean sample of cluster galaxies free of
interlopers {\em before\/} attempting a further analysis of the mass distribution in the cluster.
This final analysis can be performed via a number of methods e.g. fitting velocity dispersion
profile assuming isotropic orbits (e.g. Biviano \& Girardi 2003), fitting velocity dispersion and
kurtosis for arbitrary constant anisotropy ({\L}okas et al. 2006) etc.
The final outcome of these procedures will
depend on their specific properties and on the properties of objects to which they are applied
(e.g. whether they are spherically symmetric, depart from equilibrium, how well they are
sampled etc.). For example, Sanchis et al. (2004) and {\L}okas
et al. (2006) applied the dispersion+kurtosis fitting method to simulated clusters (after
removal of interlopers) and discussed how well the main properties of the clusters (including the
mass) are reproduced.

For the purpose of this study we used a present-day output of a pure dark matter, medium-resolution
cosmological $N$-body simulation in which cluster-size haloes can be identified. Taking advantage of
the fact that the distribution of galaxies in clusters is similar to mass distribution in simulated
dark matter haloes, i.e. both are cuspy and can be approximated by the NFW
(Navarro et al. 1997) profile (e.g. Carlberg et al. 1997; {\L}okas \& Mamon 2003;
Biviano \& Girardi 2003) we assumed that the galaxies can be approximated by just dark matter
particles. Although it would be worthwhile to test the methods on a set of higher resolution
simulations where galaxies or subhaloes can be identified, the distributions of subhaloes both
in space and velocity
are known to be biased with respect to those of dark matter particles (Diemand et al.
2004). On the other hand, Faltenbacher \& Diemand (2006) have recently shown that subhaloes with
sufficiently high mass corresponding to galaxies have distributions much less biased and very similar
to those of dark matter particles, which makes the effort of using subhaloes questionable.
Nevertheless, a possibility to assign stellar populations to subhaloes can considerably
improve their usefulness in the analysis. As shown by Biviano et al. (2006),
subhaloes with old stellar populations are more concentrated around their mother haloes
so by selecting them one can reduce the contamination by interlopers.

The paper is organized as follows. In section~2 we describe the way to create mock data sets from the
simulations, introduce different types of interlopers and discuss how the presence of interlopers
can affect the inferred properties of a galaxy cluster.
Section~3 is devoted to direct methods of interloper removal. We first discuss the dynamical
approach where the maximum velocity available to a member
galaxy is estimated using some assumptions about the cluster mass profile. Next we study the
statistical approach in its most commonly used forms which we then generalize by considering
the distribution of galaxies in projected phase space.
We also discuss the efficiency of different mass estimators in identifying interlopers.
Section~4 is devoted to indirect methods of interloper treatment and the discussion follows
in section~5.

\section{Interlopers on velocity diagrams of simulated haloes}

\begin{figure*}
\begin{center}
    \leavevmode
    \epsfxsize=17cm
    \epsfbox[50 50 700 490]{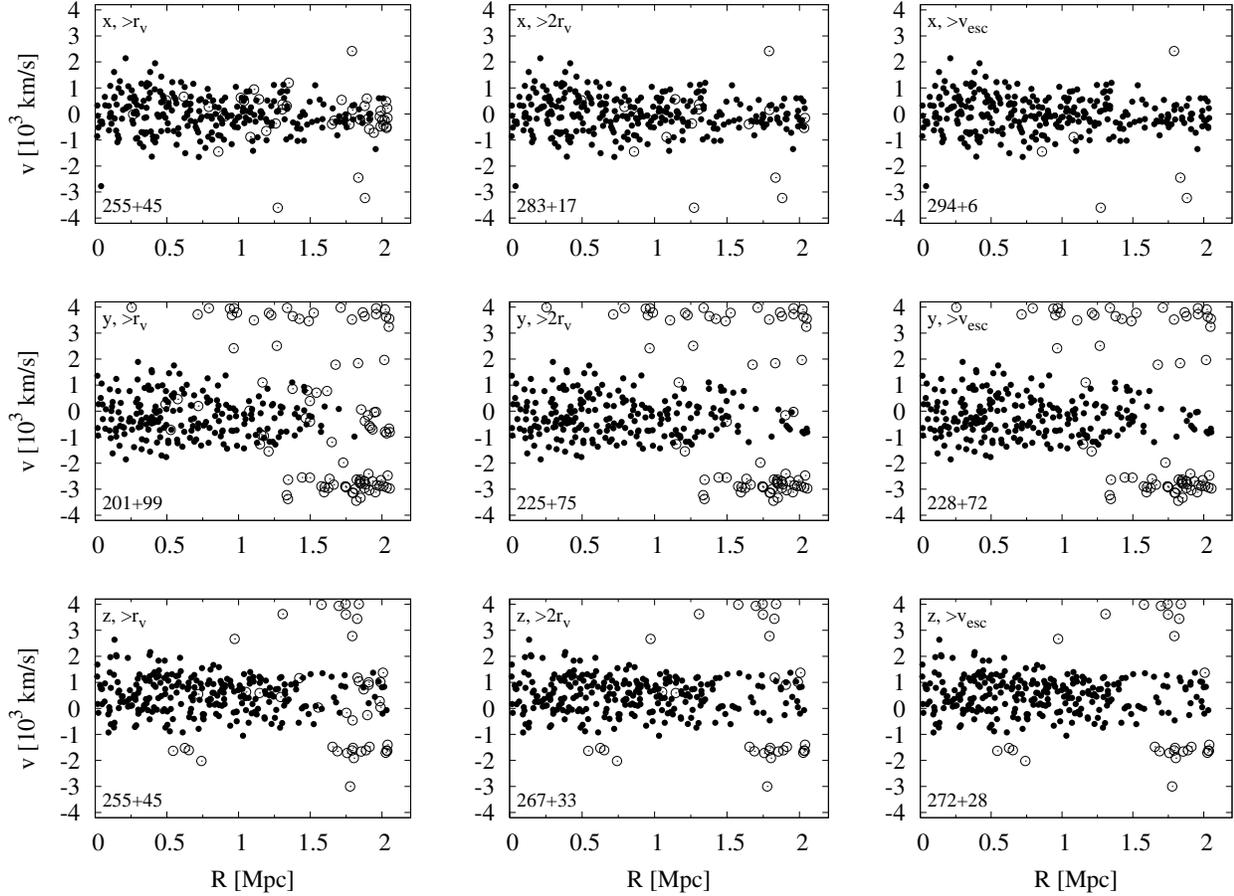}
\end{center}
\caption{Velocity diagrams of halo 6 out to $R=r_v$ in the mass centre rest frame of
reference seen in different projections and with different types of interlopers.
Filled and empty circles indicate halo particles and interlopers respectively.
In the top left corner of each panel we mark the projection axis and
the criterion for interloper identification ($>r_{v}$, $>2r_{v}$ or $>v_{\rm esc}$
for particles beyond $r_v$,
$2 r_v$ and unbound particles respectively). In the bottom left corner we give
numbers of halo particles and interlopers which are seen on a given velocity
diagram.}
\label{fig1}
\end{figure*}

In this work we used an $N$-body cosmological simulation of standard $\Lambda$CDM
model described in Wojtak et al. (2005). The simulation was performed using a version
of the ART (Adaptive Refinement Tree) code (Kravtsov et al. 1997) in a box
of size 150 $h^{-1}$ Mpc with parameters $h=0.7$, $\Omega_{M}=0.3$,
$\Omega_{\Lambda}=0.7$ and $\sigma_{8}=0.9$. From the whole sample of dark matter
haloes formed in the final simulation output ($z=0$) we choose 10 massive
(10$^{14}$-10$^{15}M_{\odot}$) and possibly relaxed ones i.e. without any obvious
signatures of ongoing major mergers. They are listed and
described in Wojtak et al. (2005). All of them are characterized by mildly
radial particle orbits and their density profiles are
well fitted up to the virial radius $r_{v}$ by the NFW formula
\begin{equation}
	\frac{\rho}{\rho_{c,0}}=\frac{\Delta_{c}c^{2}g(c)}{3(r/r_{v})(1+c(r/r_{v}))^{2}},
\end{equation}
where $g(c)=[\ln(1+c)-c/(1+c)]^{-1}$, $c$ is the concentration parameter,
$\rho_{c,0}$ is the critical density at
present and $\Delta_{c}$ is a parameter defining virial mass in terms of
overdensity with respect to the critical density. We assume
$\Delta_{c}=101.9$ which is the value valid for the concordance
$\Lambda$CDM model with $\Omega_{m}=0.3$ and $\Omega_{\Lambda}=0.7$ ({\L}okas \&
Hoffman 2001). The mean value of the concentration parameter averaged over all
10 haloes is equal to 7.2.

In order to emulate kinematic data for a galaxy cluster embedded in a given
dark mater halo we place an imaginary observer at the distance of $D$=100 Mpc
from the halo centre (going from the simulation comoving coordinates to the
observer's redshift space) so that the receding velocity of a halo mass centre
observed by him is around 7000 km s$^{-1}$. Approximating the
conical shape of the observation beam with a cylinder ($D>>r_{v}$), we  project
position vectors of simulation particles onto the plane perpendicular to the
line of sight and their velocities with respect to the observer onto his line
of sight. Assuming that some of the simulation particles represent galaxies, we
randomly select 300 particles from the inside of the observation cylinder with
projected radius $R=r_{v}$, where the virial radius $r_{v}$ is found in 3D
analysis. Additionally, we restrict
our selection to  particles with velocities from the range $\pm$ 4000 km s$^{-1}$
with respect to the velocity of a halo mass centre. This choice of velocity cut-off,
corresponding to at least $4\sigma_{\rm los}$ for cluster-size objects,
guarantees that we do not exclude any cluster galaxies with high peculiar velocities.

We place the cylinder of
observation along the main axes of the simulation box so the orientation
of the haloes (which have triaxial shapes) with respect to the line of sight
should be random. Finally for each of the 10 haloes we obtain three sets
of projected galaxy positions and velocities from observations along $x$, $y$
and $z$ axis of the simulation box. We will refer to these sets of data as
{\em velocity diagrams}. Each velocity diagram includes both particles from
the inside of a given halo (we call them simply halo particles) and particles
from the outside of a halo which are seen because of the projection effects
(we call them interlopers).

First we identify the true interlopers in our data using the full 3D
information about positions and velocities of simulation particles. Obviously
one can think about interlopers as particles which are beyond $r_{v}$
since they are from the outside of virialized region and they are not used in
the estimation of density profile. We find that on average 24 percent of
particles on our velocity diagrams have $r>r_{v}$. This criterion of interloper
identification, however, seems to be too restrictive in many cases since
the object may possess a virialized region or at least a well defined density
profile extending up to radii beyond $r_{v}$ defined by density contrast
parameter $\Delta_{c}$ (e.g. Klypin et al. 2003; Wojtak et al. 2005; Prada
et al. 2006). Besides, halo shapes are not spherical and imposing this kind
of symmetry we can lose particles that are inside the real virialized region of
the halo. We find that almost half of particles beyond $r_{v}$ on our velocity
diagrams reside below $\sim 2r_{v}$ and are bound to their halo (the fraction
of unbound particles is negligible at $2r_{v}$). These considerations suggest
that they could also be treated as good tracers of the halo potential. We have
therefore decided to consider two more conservative criteria of interloper
selection: particles beyond $2r_{v}$ and unbound particles (with velocity greater
than the escape velocity). Average contribution of these groups to the particles
on the velocity diagrams is 13 and 8 percent respectively.
Fig.~\ref{fig1} shows a set of velocity diagrams for halo 6
in different projections (rows) and with different criteria
of interloper selection (columns), where filled and empty circles correspond
to halo particles and interlopers respectively.

\begin{figure}
\begin{center}
    \leavevmode
    \epsfxsize=7.5cm
    \epsfbox[50 50 410 800]{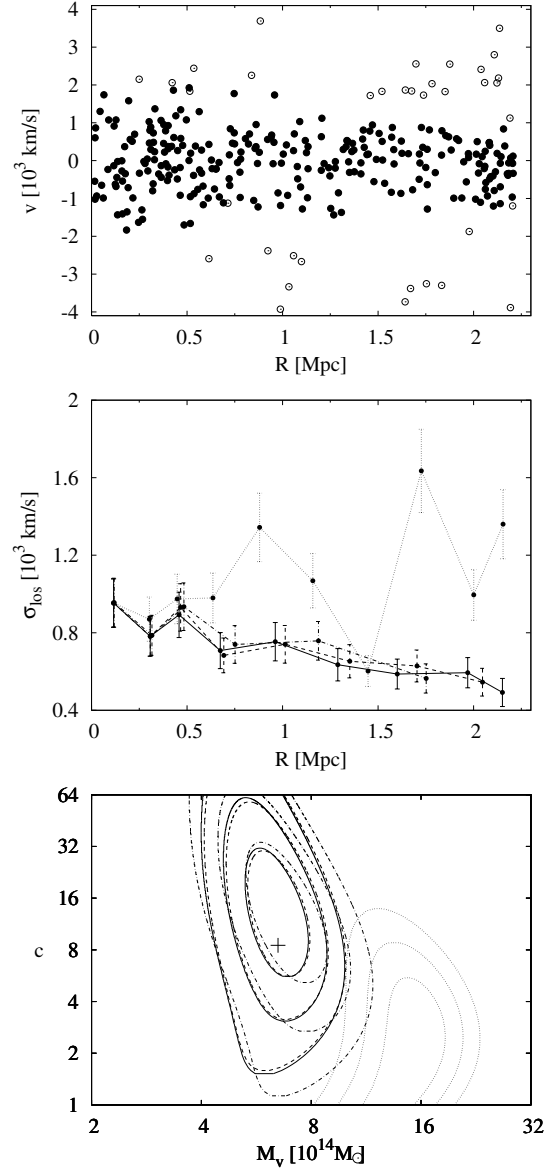}
\end{center}
\caption{The top panel shows the velocity diagram of halo 4 in projection along
$z$ axis. Filled and empty circles indicate particles bound and unbound to this
halo respectively. Middle and bottom panels show respectively the dispersion profiles and the
results of the fitting procedure in the form of 68.3, 95.4 and 99.73 percent
probability contours in the $M_v-c$ parameter plane assuming isotropic orbits.
The different types of lines correspond to different subsamples of
particles used to calculate the dispersion profile: dotted lines are for the whole
sample with interlopers included, solid ones for bound particles, dashed ones for
particles below $2r_{v}$ and dotted-dashed lines for particles below $r_{v}$.
The cross marks the concentration parameter and virial mass found in 3D analysis
of the mass distribution in the halo.}
\label{fig2}
\end{figure}

To illustrate how interlopers affect the results of dynamical analysis we fit
for simplicity an isotropic solution of the Jeans equation to a velocity dispersion
profile measured for one of the simulated velocity diagrams (see {\L}okas \&
Mamon 2001, 2003 and {\L}okas et al. 2006 for details of the Jeans formalism and the
fitting procedure). The dispersion profile is measured in radial bins for
a whole sample (300 particles) and for three subsamples of particles cleaned
of three types of interlopers introduced above. Fig.~\ref{fig2} shows a typical
velocity diagram generated from our halo 4 (with filled and empty circles
as bound and unbound particles respectively), dispersion profiles for the four
mentioned subsamples of particles and the corresponding results of the fitting
procedure aimed at estimating the virial mass $M_{v}$ and concentration $c$
of the halo, where parameter  values found in 3D analysis are marked with a cross.
All lines corresponding to the same particle sample are drawn with the lines of
the same type.

Although the results shown in Fig.~\ref{fig2} concern just a single case of a
velocity diagram, they illustrate well the general feature of bias caused by
interlopers. First, note that the velocity dispersion is overestimated mainly
in the outer part of the velocity diagram and this is caused mostly by unbound
particles (since all dispersion profiles calculated for the data cleaned of interlopers
of three different types, which include at least all unbound particles, are almost
the same). Second, all three corrected dispersion profiles infer fitting results
which are very similar to each other and include the true parameter values inside
$1\sigma$ confidence level contour. Adding unbound particles to the analysis
shifts $M_{v}$ towards higher masses (which is due to the overestimated velocity
dispersion) and forces the concentration parameter to lower values (which is due
to the rising dispersion profile).

\section{Direct methods of interloper removal}

\subsection{Overview}

In this section we study methods which allow us to remove a significant fraction of
interlopers using some criteria. First we consider restrictions on the positions of halo
particles on the
velocity diagram. Given the maximum velocity available for halo particles (dynamical
approach) or a distribution of halo particles on the velocity diagram (statistical
approach) we impose boundaries on the area of the velocity diagram likely occupied
by halo particles. Interlopers are then identified as particles from the outside
of this area. Then we consider a criterion based on the way interlopers affect different
mass estimators. In all these approaches the procedure of interloper removal is iterative.
In each step new boundaries of the area occupied by halo particles or mass estimators
are determined from the data partially cleaned of interlopers in the previous steps
and the next group of interlopers is removed. All methods are supposed to
converge after a few iterations when no more interlopers are identified.

Knowing which particles on the velocity diagrams are real interlopers (belonging to
any of the samples defined in the previous section) we are able to study
the efficiency of different methods aimed at eliminating interlopers from
velocity diagrams by comparing lists of interlopers found by these methods
with those identified in 3D analysis. To quantify these results we
introduce three parameters: a fraction of identified interlopers $f_{i}$, a
fraction of halo particles (galaxies) which were taken for interlopers by
mistake $f_{g}$ and a fraction of non-identified interlopers remaining
in the final sample of halo members $f_{h}$. For an ideal method of interloper
removal we would have all interlopers identified correctly, i.e. $f_{i}=1$,
$f_{g}=0$ and $f_{h}=0$. In order to judge the performance of different
schemes of interloper removal we
calculate the mean values of the parameters and their dispersions averaging
over the whole set of velocity diagrams.
It should be kept in mind, however, that the values of these parameters will
depend on the initial velocity cut-off used to select the data (allowing wider
velocity range we would obtain higher values of $f_{i}$). The important point
is that the relative efficiency of different methods of interloper removal
should not depend on this velocity cut-off. We address this issue further
in the last section.

\subsection{Dynamical approach}

In this approach, we identify an interloper as a particle at a given projected
radius $R$ whose velocity exceeds a maximum velocity available for halo
particles at this radius. The main problem of this method lies in the choice
of proper maximum velocity profiles. Let us consider two characteristic
velocities: the circular velocity $v_{\rm cir}$ and the infall velocity $v_{\rm inf}$
given respectively by
\begin{eqnarray}
v_{\rm cir}   & = & \sqrt{GM(r)/r}\\
v_{\rm inf}   & = & \sqrt{2}v_{\rm cir}.
\end{eqnarray}
Another quantity of interest would be the escape velocity $v_{\rm esc} = \sqrt{2|\Phi(r)|}$,
however, as discussed by Prada et al. (2003) it does not lead to any useful criterion for
interloper removal. The interpretation of the infall
velocity is as follows. Assuming circular orbits of a given set of particles one can
obviously recover the relation between potential and kinetic energy ($U$ and $T$
respectively) postulated by the virial theorem: $2T_{\rm cir}=-U_{\rm cir}$. The
infall velocity $v_{\rm inf}$ is simply an upper limit to the particles' velocities
for which the virial theorem equation is violated, $T_{\rm inf}=-U$ (den Hartog
\& Katgert 1996; Beers et al. 1982). This limit originates from the requirement
that a given particle is bound to the halo ($U+T<0$).
Note that this condition provides a stronger restriction on the maximum velocity
than the general formula for the escape velocity since $v_{\rm
cir}<\sqrt{|\Phi(r)|}$. The equality $v_{\rm esc}=v_{\rm inf}$ would only occur if
the density distribution dropped to zero at $r$ since then we would have $|\Phi(r)| =
GM(r)/r$. The velocity $v_{\rm inf}$ can therefore be viewed as an escape velocity from
the mass interior to $r$.

Following den Hartog \& Katgert (1996) we introduce two formulae for the
maximum velocity profile. First, assuming that the direction of particle velocity in the
limit determined by $v_{\rm inf}$ has any orientation, the maximum velocity at
a given projected radius $R$ is given by
\begin{equation}\label{vmax_inf}
	v_{\rm max}={\rm max}_{R}\{v_{\rm inf}\},
\end{equation}
where ${\rm max}_{R}$ is a maximum along the line of sight at the distance $R$
from the halo centre. A second, more restrictive criterion which gives more
accurate limits at high $R\sim r_{v}$ can be obtained from
\begin{equation}	\label{vmax_cir}
	v_{\rm max}={\rm max}_{R}\{v_{\rm inf} \cos\theta, v_{\rm cir} \sin\theta \} ,
\end{equation}
where $\theta$ is the angle between position vector of the particle with respect
to the halo centre and the line of sight. With this formula, we assume
a special kinematic model which allows particles to fall onto the halo centre
with velocity $v_{\rm inf}$ or to move in a tangential direction with circular
velocity $v_{\rm cir}$.

\begin{figure}
\begin{center}
    \leavevmode
    \epsfxsize=8cm
    \epsfbox[50 50 400 300]{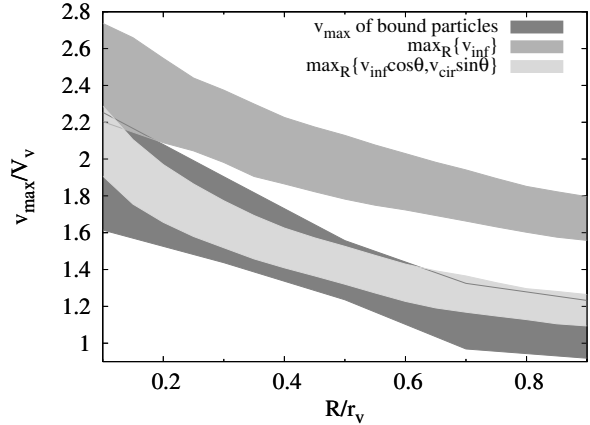}
\end{center}
\caption{Maximum velocity profiles. The dark gray area in the background indicates
maximum velocity reached by bound particles. Medium and light gray strips correspond
respectively to the formula (\ref{vmax_inf}) and (\ref{vmax_cir}) with the mass profile
given by the mass estimator (\ref{M_VT}). Widths of the shaded areas are given by the
dispersions following from averaging over the whole sample of velocity diagrams.
}
\label{fig3}
\end{figure}

To complete the above prescription for the maximum velocity profiles one needs to specify the
mass profile. As proposed by den Hartog \& Katgert (1996), we use the mass estimator
$M_{VT}$ derived from the virial theorem (Limber \& Mathews 1960; Bahcall \&
Tremaine 1981; Heisler et al. 1985)
\begin{equation}	\label{M_VT}
	M_{VT}(r=R_{\rm max})=\frac{3\pi N}{2G}\frac{\Sigma_{i}
	(v_{i}-\bar{v})^{2}}{\Sigma_{i<j}1/R_{i,j}} ,
\end{equation}
where $N$ is a number of galaxies enclosed on the sky by a circle
with radius $R_{\rm max}$, $v_{i}$ is the velocity of the $i$-th galaxy and
$R_{i,j}$ is a projected distance between $i$-th and $j$-th galaxy. This formula
is valid for spherical systems with arbitrary anisotropy.
The mass profile can be simply obtained as $M(r)\approx
M_{VT}(R_{i}<r<R_{i+1})$, where $R_{i}$ is the sequence of projected radii of
particles (galaxies) in the increasing order. The virial theorem applies to a whole
system and otherwise one needs to add a surface term (e.g. The \& White 1986).
Recently Biviano et al. (2006) estimated its value for simulated clusters within an
aperture of 1.5 $h^{-1}$ Mpc. However,
since our purpose is not to estimate accurately the mass but to design a procedure
for interloper removal, the formula (\ref{M_VT}) is sufficient.

\begin{figure}
\begin{center}
    \leavevmode
    \epsfxsize=8cm
    \epsfbox[50 50 410 550]{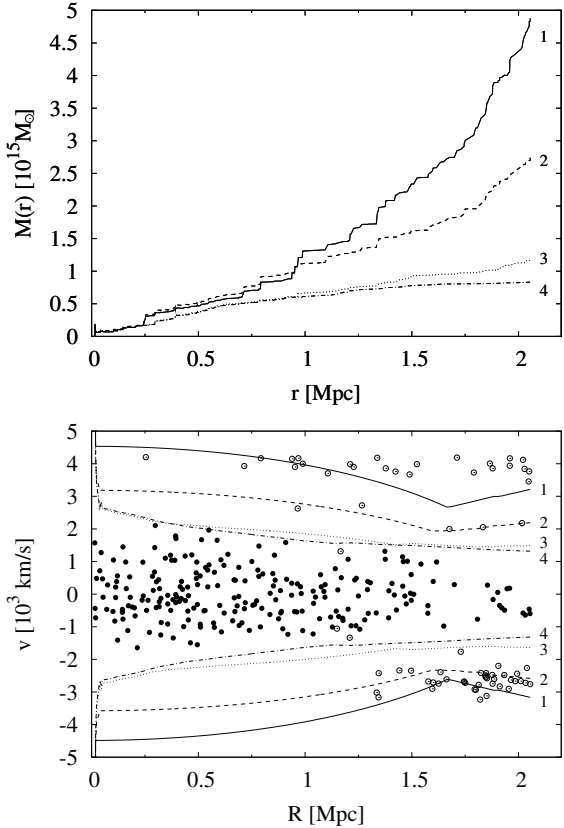}
\end{center}
\caption{Illustration of successive steps in $v_{\rm max}(1)$ method of interloper
removal for halo 6 in projection along the $y$ axis. The method uses
the maximum velocity profile (\ref{vmax_cir}) and mass profile
(\ref{M_VT}). The top and bottom panels show successive mass profiles and
maximum velocity profiles separating interlopers from halo particles respectively.
Filled and empty circles mark particles bound and unbound to the halo. Numbers
indicate successive steps of the procedure which are described by the lines of
different types. Final maximum velocity profiles (number 4) are drawn with
dashed-dotted lines.}
\label{fig4}
\end{figure}

Using subsamples of bound particles gathered on all mock velocity diagrams
we calculate the maximum velocity profiles given by (\ref{vmax_inf}) and
(\ref{vmax_cir}) with the mass profile determined by (\ref{M_VT}). The results,
expressed in units of circular velocity at the virial radius $V_v$,
are shown in Fig.~\ref{fig3} as medium and light gray strips respectively,
whereas the dark gray profile seen in the background of this plot
indicates the average maximum velocity reached by bound particles
(including bound particles beyond $r_{v}$) on any velocity diagram.
Widths of all three areas are given by the dispersions resulting from averaging
the maximum velocity profiles over the whole sample of velocity diagrams.
It is clear that formula (\ref{vmax_cir}) is expected to work best in removing interlopers
since its profile (light gray) coincides almost exactly with the maximum velocity
reached by bound particles (dark gray). On the other hand, the profile generated 
by formula (\ref{vmax_inf}) seems too conservative to be useful.

Fig.~\ref{fig4} illustrates successive steps of interloper removal with the maximum velocity
(\ref{vmax_cir})  (hereafter $v_{\rm max}(1)$ method) for one of mock velocity diagrams with
rather large number of unbound particles. The top and bottom panels show mass
profiles and maximum velocity profiles separating interlopers from halo particles on the
velocity diagram for successive iterations of this method marked with numbers. The final virial
mass given simply by the value of estimator $M_{VT}$ for $R_{\rm max}=r_{v}$ is equal to
$8.35 \times 10^{14}M_{\odot}$ which is a few times lower than for the total contaminated sample
(first iteration in Fig.~\ref{fig4}) and reasonably close to the real value of
the virial mass found in 3D analysis, $5.35 \times 10^{14}M_{\odot}$. Note that
the mass estimator $M_{VT}$ is known to overestimate the true mass due to
the neglect of the surface term (see Biviano et al. 2006) and a more reliable
final estimate can in general be obtained by fitting velocity moments (Sanchis et
al. 2004; {\L}okas et al. 2006).

\begin{table}
\caption{Results of different procedures of interloper removal
in terms of the fraction of removed interlopers $f_{i}$, the fraction of
halo particles incorrectly identified as interlopers $f_{g}$ and the
fraction of non-identified interlopers remaining in the final sample
of halo members $f_{h}$. The table lists both the mean values
$\langle f\rangle$ and the dispersions $\sigma_{f}$ of the
parameters obtained in the analysis of 30 mock velocity diagrams.
For all seven methods of interloper removal considered in
section 3 results are quantified for three
different definitions of interlopers, particles beyond $r_{v}$ or $2r_{v}$
and unbound particles, as marked in the second column (interloper
type -- i/t) by $r_{v}$, $2r_{v}$ and $v_{\rm esc}$ respectively.
}
\label{methods}
\begin{center}
\begin{tabular}{llcccc r@{.}l c}
method & i/t &$\langle f_{i}\rangle$ & $\sigma_{f_{i}}$ & $\langle f_{g}\rangle$ &
$\sigma_{f_{g}}$ & \multicolumn{2}{c}{$\langle f_{h}\rangle$}& $\sigma_{f_{h}}$ \\
\hline
\hline
$v_{\rm max}(1)$ & $r_{v}$    & 23 & 17 & 1.1 & 1.5 &  19&7   & 8.7 \\
         & $2r_{v}$  & 48 & 26 & 1.1 & 1.4 &    7&7   & 7.8 \\
         & $v_{\rm esc}$& 73 & 23 & 1.0 & 1.4 &    2&4   & 4.0 \\
\hline
$v_{\rm max}(2)$ & $r_{v}$    & 13 & 11 & 0.0 & 0.0 &  21&9   & 9.4 \\
         & $2r_{v}$  & 30 & 22 & 0.0 & 0.0 &  10&2   & 9.0 \\
         & $v_{\rm esc}$& 48 & 29 & 0.0 & 0.0 &    5&2   & 6.5 \\
\hline
$3\sigma_{\rm los}(5)$   & $r_{v}$    & 17 & 9   & 0.9 & 1.4 &  21&3   & 8.9 \\
         & $2r_{v}$  & 37 & 21 & 0.8 & 1.2 &  9&4     & 8.3 \\
         & $v_{\rm esc}$& 58 & 27 & 0.7 & 1.0 &    4&3   & 5.3 \\
\hline
$3\sigma_{\rm los}(10)$ & $r_{v}$    & 19 & 10 & 1.6 & 2.4 &  21&0   & 8.8 \\
         & $2r_{v}$  & 40 & 22 & 1.5 & 2.4 &    9&1   & 8.3 \\
         & $v_{\rm esc}$& 63 & 27 & 1.4 & 2.2 &    3&9   & 5.1 \\
\hline
$3\sigma_{\rm los}(R)$   & $r_{v}$    & 19 & 17 & 0.2 & 0.5 &  20&4   & 9.2 \\
                 & $2r_{v}$  & 40 & 25 & 0.2 & 0.4 &    8&6   & 8.4 \\
                 & $v_{\rm esc}$& 61 & 28 & 0.2 & 0.3 &    3&4   & 4.4 \\
\hline
$v_{\rm lim}$      & $r_{v}$    & 19 & 17 & 0.3 & 0.5 &  20&4   & 9.2 \\
                 & $2r_{v}$  & 41 & 25 & 0.3 & 0.5 &    8&5   & 8.3 \\
                 & $v_{\rm esc}$& 62 & 27 & 0.3 & 0.4 &    3&4   & 4.4 \\
\hline
$M_{P}/M_{VT}$         & $r_{v}$    & 18 & 7  & 1.2 & 1.2 &  21&1   & 8.9 \\
                 & $2r_{v}$  & 40 & 22 & 1.2 & 1.3 &    9&3   & 8.3 \\
                 & $v_{\rm esc}$& 65 & 26 & 1.2 & 1.2 &    4&1   & 5.8 \\
\hline
\hline
\end{tabular}
\end{center}
\end{table}

With this method, on average 73 percent of unbound particles are identified
and removed from a sample and only around 1 percent of bound particles are taken
for interlopers and lost from the velocity diagram so that the final samples include
only around 2-3 percent of unbound particles (see Table~\ref{methods} for details).
Note that the fraction of removed interlopers $f_i$ is limited in principle to values
lower than about 75 percent because roughly $1/4$ of unbound particles within the
observation cylinder with velocity cut-off 4000 km s$^{-1}$ are within the envelope of
bound velocities and therefore inaccessible for direct methods of interloper identification.
Since $f_{i}=73$ percent available in this approach is very close to the expected maximum,
the method presented above is possibly the most effective. As expected from
Fig.~\ref{fig3}, the method of interloper removal based on profile (\ref{vmax_inf}) is
too conservative and on average identifies much less interlopers
than the previous one (see $v_{\rm max}(2)$ method in Table~\ref{methods}).

\subsection{Statistical approach}

The idea of this approach is to use the information about the distribution of
halo particles on the velocity diagram to distinguish between the probable
halo particles and interlopers. The first scheme along these lines was introduced
by Yahil \& Vidal (1977) who proposed to identify interlopers as galaxies with
velocities from the outside of the range $\pm3\sigma_{\rm los}$ around the
mean cluster velocity, where $\sigma_{\rm los}$ is the projected velocity dispersion
of galaxies in the cluster given by the standard unbiased estimator. In this formulation
the method is model-independent so that the data are self-verified as far as the
interloper removal is concerned.

It is easy to generalize the above prescription to the case of data gathered in
$n$ radial bins so that $3\sigma_{\rm los}$ procedure could be applied in each
bin independently in the way proposed by Yahil \& Vidal. This modification allows
us to take into account dependence of the velocity dispersion on $R$. However, increasing
the number of bins we let the dispersion in the outer part of velocity diagram be much more
overestimated by interlopers. A way to overcome this problem is to
use subsequently different numbers of bins. The dispersion in
wide bins (when there is a small number of bins) is less biased by
interlopers so in this case we remove interlopers efficiently. On the
other hand, using narrow bins (when there is a larger number of bins) we
measure the dispersion locally taking into account the dependence of
$\sigma_{\rm los}$ on $R$.

In each step of this method we use the following estimators of mean velocity
and velocity dispersion
\begin{eqnarray}
	\bar{v}_{i} & = & \frac{\Sigma_{j=1}^{j=m-1}v_{i,j}}{m-1}\\
	\sigma^{2}_{{\rm los},i} & = & \frac{\Sigma_{j=1}^{j=m-1}(v_{i,j}-
	\bar{v}_{i})^{2}}{m-2},
\end{eqnarray}
where $i$ is a number of bin, $m$ is a number of datapoints per bin and $v_{i,j}$ is
the sequence of velocities in the $i$-th bin with the most outlying from the mean
value in the last position so that following the prescription of Yahil \& Vidal
we do not take into account these velocities in estimating the dispersion.
For each number of bins changing in a given step from $n_{\rm min}$ to
$n_{\rm max}$ we remove particles with velocities from the outside of the range
$\pm 3\sigma_{{\rm los},i}$ around $\bar{v}_{i}$:
$|v_{i,j}-\bar{v}_{i}|>3\sigma_{{\rm los},i}$. The procedure converges after a few
steps when no more interlopers are found in any bin.

\begin{figure}
\begin{center}
    \leavevmode
    \epsfxsize=8.5cm
    \epsfbox[50 50 770 300]{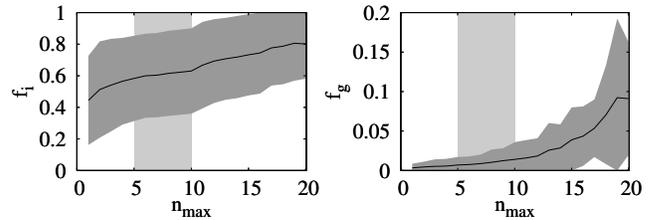}
\end{center}
\caption{
Dependence of the results of the interloper removal method based on the
analysis of the velocity dispersion of binned data on the maximum number
of bins $n_{\rm max}$. The left panel shows the fraction of identified unbound particles
$f_{i}$ and the right one the fraction of bound particles taken for interlopers
by mistake $f_g$. Widths of both profiles (dark gray) are given by the dispersions following
from averaging over the sample of velocity diagrams. Light gray strip in the background
indicates the best range of $n_{\rm max}$.
}
\label{fig8}
\end{figure}

\begin{figure}
\begin{center}
    \leavevmode
    \epsfxsize=8.5cm
    \epsfbox[50 50 590 610]{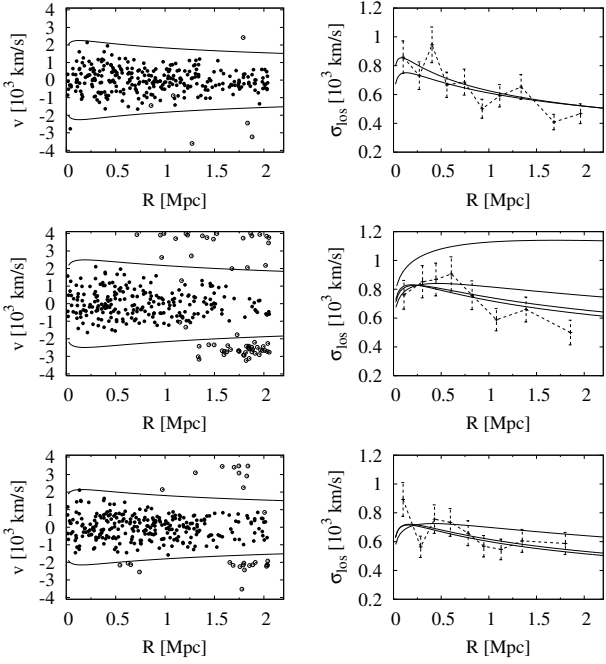}
\end{center}
\caption{Results of the $3\sigma_{\rm los}(R)$ method for halo 6 in three projections
($x$, $y$ and $z$ from top to bottom). Left column panels show velocity
diagrams with final $\pm 3\sigma_{\rm los}(R)$ lines separating interlopers
from halo particles. Filled and empty circles on velocity diagrams indicate
bound and unbound particles respectively. Right column panels show the line-of-sight
velocity dispersion
profiles obtained in successive steps of the procedure (solid lines from top to
bottom). The dashed line with error bars is the dispersion profile measured for
bound particles. }
\label{fig7}
\end{figure}

We find that in order to remove even strongly clustered groups of interlopers
(like the ones on the velocity diagram of halo 6 in projection along $y$ axis
seen in the middle left panel of Fig.~\ref{fig7}) it is necessary to fix $n_{\rm min}=1$
which corresponds to the original approach of Yahil \& Vidal (1977). To fix a
maximum number of bins $n_{\rm max}$ we consider the dependence of $f_{i}$ and
$f_{g}$ on different choices of the value of $n_{\rm max}$. The results are
shown in Fig.~\ref{fig8} in the form of dark gray profiles. Using them
we can easily find the values of $n_{\rm max}$ for which
the procedure gives possibly high $f_{i}$ and low $f_{g}$. This range of
$n_{\rm max}$ is marked with a light gray strip in the background of the plot.
Applying its lower limit ($n_{\rm max}=5$), for which $f_{i}$ profile begins inclining
towards $n_{\rm max}$ axis, leads to slightly conservative method of interloper removal
with average rate of unbound particle identification $f_{i}=58$ percent
(see $3\sigma_{\rm los}(5)$ method
in Table~\ref{methods}). In the upper limit of this range ($n_{\rm max}=10$), when
$f_{g}$ starts increasing rapidly, algorithm is a bit more restrictive and allows to remove on
average 63 percent of unbound particles with the rate of misidentification $f_g$
comparable to the result of $v_{\rm max}(1)$ method (see $3\sigma_{\rm los}(10)$
method in Table~\ref{methods}
for more details).

Recently {\L}okas et al. (2006) generalized the $\pm3\sigma_{\rm los}$ rule of interloper
identification to a continuous velocity dispersion profile: $\pm3\sigma_{\rm los}(R)$, where
$\sigma_{\rm los}(R)$ is the projected isotropic solution to Jeans equation parametrized by
$M_{v}$ and $c$ (Binney \& Mamon 1982; see also Prugniel \& Simien 1997; {\L}okas \&
Mamon 2001; Mamon \& {\L}okas 2005) fitted to the binned data. Assumption of
isotropic orbits allows us to break the degeneracy between $c$ and the anisotropy
and to trace accurately the shape of the velocity dispersion profile with the $c$
parameter only. We find that in some cases
of velocity diagrams with strong interloper contamination this procedure stops too early
because of the overestimation of the velocity dispersion profile. To fix this problem we
propose to fit $\sigma_{\rm los}(R)$ to an incomplete dispersion profile after rejecting a few
outer data points which are most contaminated. In our case we proceeded with the fitting
for at least 6 data points to the maximum of 10 (always with 30 particles per bin) and
then used the mean values of $M_{v}$ and $c$ obtained for $k=6-10$ data points.
The mean values were weighted with the goodness of fit measure $\chi^{2}_{\rm min}/(k-2)$
so that parameter values coming from worse-quality fits caused mainly by the presence
of interlopers were naturally attenuated.

Fig.~\ref{fig7} illustrates this approach both in the form of the final
$\pm 3\sigma_{\rm los}(R)$ lines on the velocity diagram (left column) and the
velocity dispersion profiles obtained in subsequent steps of this procedure
(right column) for halo 6 in three projections. The procedure allows to remove
on average 61 percent of unbound particles from a given velocity diagram
with the rate of misidentification of only around 0.2 percent (see $3\sigma_{\rm los}(R)$ method in
Table~\ref{methods}). Note that $f_{i}$ in this case achieves values similar to those obtained
with the $3\sigma_{\rm los}(10)$ method, while $f_{g}$ is much smaller.

\subsection{Generalized statistical approach}

The methods presented in the above subsection assume implicitly that the projected
distribution of halo particles does not depend on the projected radius. In the following,
we reformulate this approach properly, taking into account the full dependence of the
distribution of halo particles on $R$ and $v$. A natural extension of the criterion
introduced by Yahil \& Vidal (1977) is then given by the boundary line $\pm v_{\rm lim}(R)$
which determines an area occupied by halo particles on the velocity diagram with
some probability $p_{\rm lim}(R)$. Conditions for $v_{\rm lim}(R)$ can be written
as follows:
\begin{eqnarray}
	\int_{0}^{R_{\rm max}}\int_{-v_{\rm lim}(R)}^{v_{\rm lim}(R)}p_{R,v}
	\textrm{d}v\textrm{d}R & = & p_{\rm lim}   \label{v_lim_eq} \\
	p_{R,v}[R,v_{\rm lim}(R)]\textrm{d}v\textrm{d}R &
	= & C\textrm{d}v\textrm{d}R   \nonumber
\end{eqnarray}
where $p_{R,v}$ is the projected probability distribution of halo particles and $C$ is its
constant value along the boundary line $\pm v_{\rm lim}$. Second equation in (\ref{v_lim_eq})
is necessary to fully constrain the final solution.

\begin{figure}
\begin{center}
    \leavevmode
    \epsfxsize=8cm
    \epsfbox[50 50 400 300]{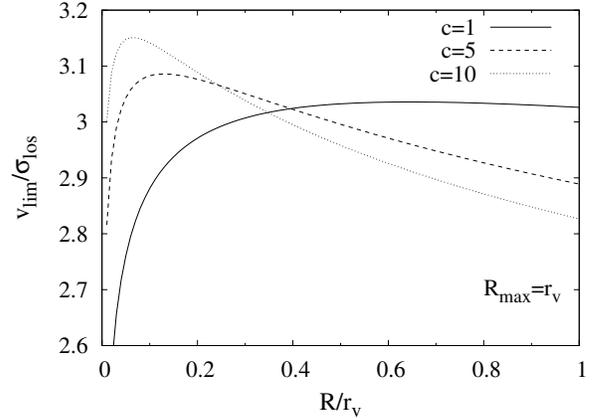}
\end{center}
\caption{Boundary lines $v_{\rm lim}$ of the area occupied by halo particles with
probability $p_{\rm lim}=0.9973$ for three values of the concentration parameter
$c=1,5,10$. The velocity diagram is assumed to have a cut-off
in radius at $R_{\rm max}=r_{v}$ and $v_{\rm lim}$ is expressed in units of the
local value of the projected velocity dispersion $\sigma_{\rm los}(R)$.}
\label{fig6}
\end{figure}

In order to find a useful analytical approximation for $p_{R,v}$ we first
assume that the probability of finding a
halo particle inside an infinitesimal range of radius $[R,R+\textrm{d}R]$ of a cylinder of
observation with radius $R_{\rm max}$ is given by
\begin{equation}
	p_{R}\textrm{d}R=2\pi
	R\frac{\Sigma_{M}(R,c)}{M_{P}(R_{\rm max},c)}\textrm{d}R, \label{p_R}
\end{equation}
where $\Sigma_{M}(R,c)$ and $M_{P}(R_{\rm max},c)$ are surface density and projected
mass inferred from this surface density; in our case they follow from the NFW density formula
(see Bartelmann 1996; {\L}okas \& Mamon 2001). To be precise, equation (\ref{p_R})
is satisfied best by dark matter particles and could be less suitable for description
of spatial distribution of galaxies. However, cluster data are consistent with galaxy distribution
being given by the NFW profile and mass-to-number density being constant within the virial radius
(see e.g. Biviano \& Girardi 2003) so that formula (\ref{p_R}) seems to be useful also
in the context of real data.

Second, we assume that the distribution of the
line-of-sight velocity at a given radius $R$ (i.e. the conditional probability of a particle
having a velocity $v$ if it is at projected distance $R$) can be well approximated by a
Gaussian distribution
\begin{equation}        \label{p_v}
	p_{v}\textrm{d}v=\frac{1}{\sqrt{2\pi}\sigma_{\rm los}(R,c)}
	\exp\Big(-\frac{v^{2}}{2\sigma^{2}_{\rm los}(R,c)}\Big)\textrm{d}v ,
\end{equation}
where $\sigma_{\rm los}(R)$ is the projected isotropic solution to the Jeans equation.
Although this approximation is not exactly valid since departures from Gaussianity are
seen both in simulated haloes and real clusters (Kazantzidis et al. 2004;
Wojtak et al. 2005; Hansen et al. 2006; {\L}okas et al. 2006) we claim it is
sufficient for our needs. In particular, one can show that for isotropic orbits the projected
velocity distribution following from the distribution function for the NFW profile (see
{\L}okas \& Mamon 2001; Widrow 2000) is remarkably close to a Gaussian in a wide
range of radii.

Finally, combining the formulae for $p_{R}$ and $p_{v}$ we get a heuristic
expression for the projected probability distribution:
\begin{equation}	\label{p_Mv}
	p_{R,v}\textrm{d}R\textrm{d}v=p_{R}p_{v}\textrm{d}R\textrm{d}v.
\end{equation}
One can immediately check that the normalization condition on the available area of the
velocity diagram is satisfied automatically with sufficiently high numerical precision since
$v_{\rm max}(R)\gtrsim 4\sigma_{\rm los}(R)$:
\begin{equation}
	\int_{0}^{R_{\rm max}}\int_{-v_{\rm max}(R)}^{v_{\rm max}(R)}
	p_{Rv}\textrm{d}v\textrm{d}R=1.
\end{equation}
Note that $p_{R,v}$ given by (\ref{p_Mv}), by analogy with results obtained by
Maoz \& Bekenstein (1990), maximizes Shannon's entropy for known functions $\Sigma_{M}(R)$
and $\sigma_{\rm los}(R)$. In the language of the information theory this means
that $p_{R,v}$ is the most plausible probability distribution given $\Sigma_{M}(R)$ and
$\sigma_{\rm los}(R)$.

Substituting (\ref{p_Mv}) into (\ref{v_lim_eq}) we derive $v_{\rm lim}(R)/\sigma_{\rm los}(R)$
from the second of these equations. Introducing this expression to the first equation
we calculate numerically the constant $C$ and in the end evaluate the whole
profile $v_{\rm lim}(R)/\sigma_{\rm los}(R)$. Fig.~\ref{fig6} shows numerical
solutions assuming cut-off radius $R_{\rm max}=r_{v}$ and $p_{\rm lim}=0.9973$, a value
corresponding to the $\pm 3 \sigma$ range for a Gaussian distribution.
We see that in general $v_{\rm lim}(R)/\sigma_{\rm los}(R)$ departs significantly from the value
of 3. Only in the limit of $p_{R}\sim \textrm{const}$ we recover the continuous version of the criterion
introduced by Yahil \& Vidal (1977): $v_{\rm lim}(R)=3\sigma_{\rm los}(R)$. We find that using the
exact solution to the set of equations (\ref{v_lim_eq}) does not improve the performance of the method,
the numbers of interlopers removed in this case are similar as for the simpler $3 \sigma_{\rm los}$
schemes (see $v_{\rm lim}$ method in Table~\ref{methods}).

\subsection{Mass estimators as indicators of interloper fraction}

\begin{figure}
\begin{center}
    \leavevmode
    \epsfxsize=8.5cm
    \epsfbox[50 50 750 550]{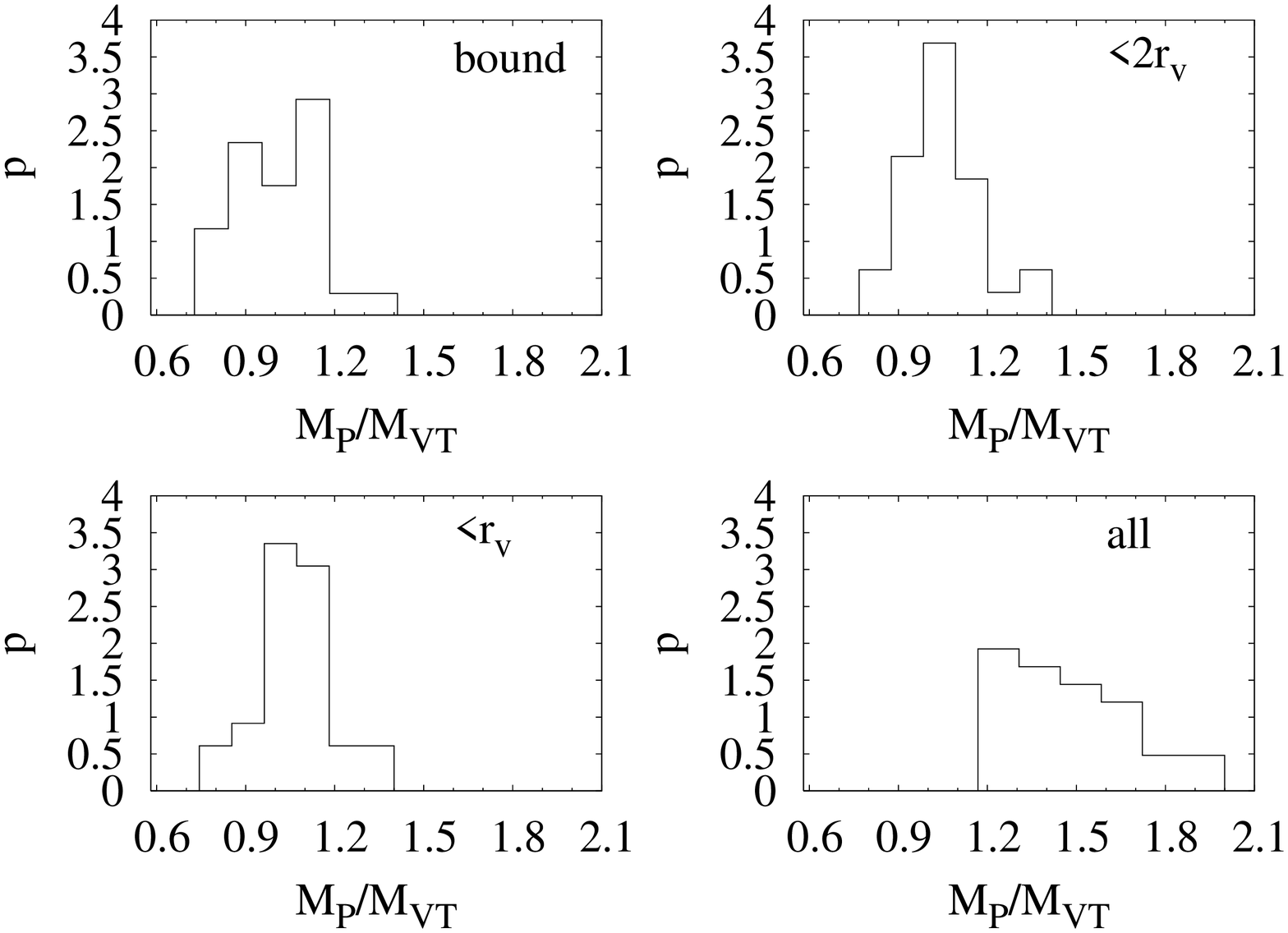}
\end{center}
\caption{Histograms of the ratio $M_{P}/M_{VT}$ for different subsamples
of particles. Subsamples are identified in the upper right corner
of each panel: the lower right panel is for all particles, upper left for
bound particles and the remaining two for particles inside $r_v$ or $2 r_v$.
All histograms are normalized to unity.}
\label{fig10}
\end{figure}

In this subsection we study the effect of interlopers on the values of two standard mass
estimators and use the results to construct another method of interloper removal. The estimators
we use are the virial mass $M_{VT}$ expressed by formula (\ref{M_VT}) and the projected
mass $M_{P}$ for isotropic orbits given by (Heisler et al. 1985; Perea et al. 1990)
\begin{equation}	\label{M_P}
	M_{P}=\frac{32}{\pi GN}\Sigma_{i}(v_{i}
	-\bar{v})^{2}R_{i}
\end{equation}
where $v_{i}$ and $R_{i}$ are the velocity and projected radius of the $i$-th
galaxy, $N$ is the number of galaxies in the sample.

Both mass estimators are sensitive to the presence of interlopers,
but each of them in a different way (Perea et al. 1990). $M_{P}$
is considerably more overestimated than $M_{VT}$ so that interlopers
effectively give rise to the increase of
$M_{P}/M_{VT}$ ratio above 1. In principle, it is impossible to relate the
value of this ratio to the number of any kind of interlopers because of
strong degeneracy: a given value of $M_{P}/M_{VT}$ can be reproduced
with various numbers of interlopers and their different distributions on the
velocity diagram. Nevertheless, it is interesting to study the distributions of the
ratio $M_{P}/M_{VT}$ for four different particle subsamples: all particles,
particles inside $2r_{v}$ or $r_{v}$ and bound particles.

The results are shown in Fig.~\ref{fig10} in the form of histograms
(normalized to unity) constructed from the data in our 30 velocity diagrams.
As we can see, there is no significant
difference between histograms obtained for all three subsamples with
different types of interlopers subtracted. All of them have a maximum at around
$M_{P}/M_{VT}=1$ and a spread between $\sim 0.8$ and $\sim1.4$. Including
unbound particles in the analysis, we get a highly asymmetric histogram, shifted
to the range $1.2-2.0$ with a maximum at $\sim 1.2$. These results prove that
the unbound particles are the ones that give rise to the overestimation of the
ratio of mass estimators and therefore the ratio $M_{P}/M_{VT}$ can
be a useful indicator of the contamination of a sample with unbound particles.

\begin{figure}
\begin{center}
    \leavevmode
    \epsfxsize=8.5cm
    \epsfbox[50 70 750 550]{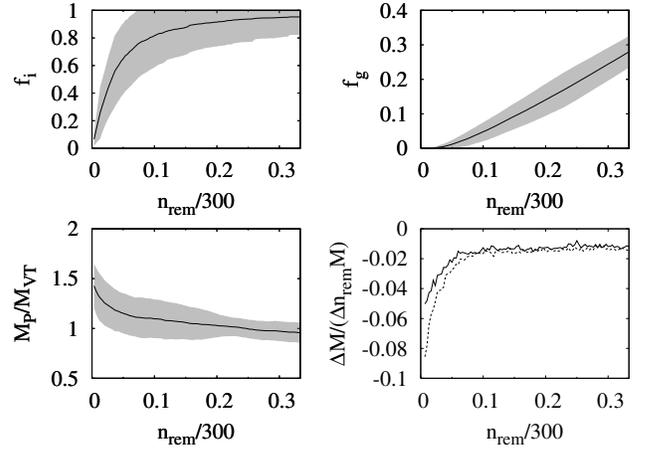}
\end{center}
\caption{The ratio of mass estimators $M_{P}$ and $M_{VT}$, their derivatives
$\Delta M/(\Delta n_{\rm rem}M)$ and fractions of unbound $f_{i}$ and bound $f_{g}$
particles removed in the procedure based on jackknife statistics as a function
of the total number of removed particles $n_{\rm rem}$. All profiles are averaged
over the whole sample of 30 velocity diagrams and the shaded areas indicate
the dispersion of values. The solid and dashed lines in the lower right
panel correspond respectively to $M_{VT}$ and $M_{P}$.}
\label{fig11}
\end{figure}

\begin{figure*}
\begin{center}
    \leavevmode
    \epsfxsize=17cm
    \epsfbox[50 50 1120 300]{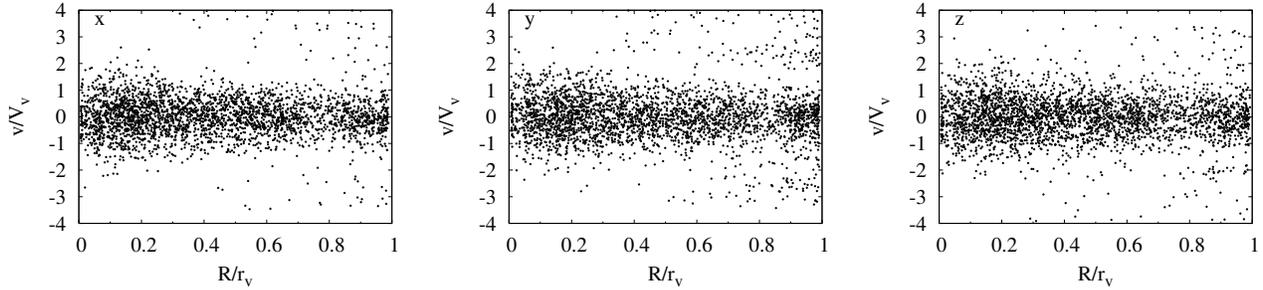}
\end{center}
\caption{Stacked velocity diagrams in projection along $x$, $y$ and $z$ axes.}
\label{fig12}
\end{figure*}

This phenomenological relation between the ratio of mass estimators and the
presence of unbound particles in the sample can be used to construct a
procedure which eliminates this kind of interlopers
from the velocity diagram. The prescription for such interloper identification
was proposed by Perea et al. (1990) and is based on jackknife
statistics. Let $\{R_{i},v_{i}\}$ be a sequence of data, where $i$ goes from 1
to $n$. Following the jackknife technique, we calculate $n$ values of both mass
estimators which correspond to $n$ subsequences with one data point excluded.
Finally we identify as an interloper the particle for which the corresponding
subsequence is the source of the most discrepant value of one of the estimators
with respect to the mean value. In the next step, the same procedure is applied
to a new data set with $n-1$ particles.

The main problem of this procedure lies in defining properly the condition for stopping
the algorithm. In order to specify it we calculate the  ratio $M_{P}/M_{VT}$ and the
fractions $f_{i}$ and $f_{g}$ (for interlopers defined as unbound particles) as functions
of the number of removed particles $n_{\rm rem}$ determined by the jackknife
technique. Fig.~\ref{fig11} shows the results averaged over all 30 velocity diagrams
so that the shaded areas indicate the dispersion of the values. From the behaviour
of all profiles we infer that the most accurate moment of convergence (around
$n_{\rm rem}/300=0.05$) coincides clearly with a characteristic knee-like point
of $M_{P}/M_{VT}$ and $\Delta M/M$ profiles. This point is usually more recognizable
in the case of single velocity diagrams so that it can be effectively used to judge where
the algorithm should be stopped (see Wojtak \& {\L}okas 2006).
Taking $n_{\rm rem}/300=0.05$ we are able
to remove on average 65 percent of unbound particles with around 1 percent of
bound particles taken for interlopers by mistake (see $M_{P}/M_{VT}$ method in
Table~\ref{methods}).

\section{Indirect methods of interloper treatment}

\subsection{The distribution of interlopers}

The key idea of the indirect approach is to treat interlopers
statistically in the proper dynamical analysis so that no particles need to be rejected
(van der Marel et al. 2000; Prada et al. 2003). In all versions of this method one
assumes that the probability distribution $p(R,v)$ of particles seen on the velocity
diagram consists of two terms which describe the distribution of halo particles and
the distribution of interlopers.

\begin{figure}
\begin{center}
    \leavevmode
    \epsfxsize=8.5cm
    \epsfbox[50 50 480 500]{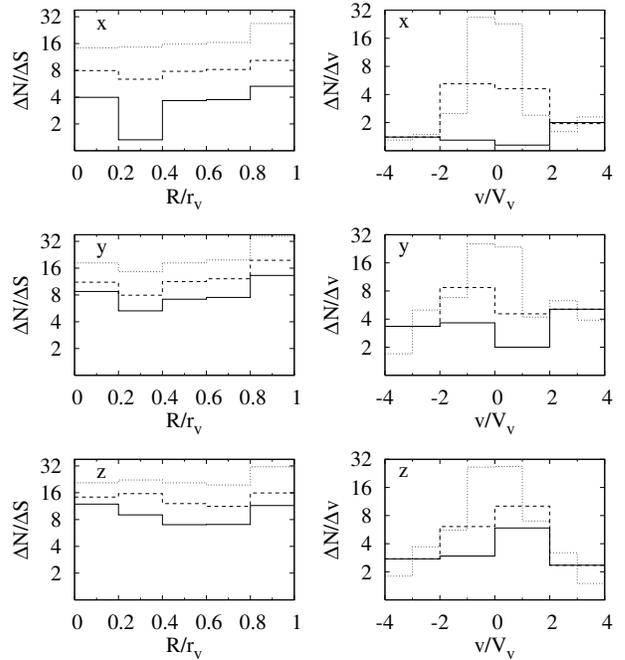}
\end{center}
\caption{Surface density (left column) and velocity histograms (right column) of
interlopers on stacked velocity diagrams in different projections marked in the
upper left corner of each panel. Solid, dashed and dotted lines correspond to
unbound particles and particles beyond $2r_{v}$ and $r_{v}$ respectively. Both
surface density and velocity histograms are normalized to the number of a given
type of interlopers averaged over 10 velocity diagrams in a given projection.}
\label{fig13}
\end{figure}

Such an analysis requires numerous and regular samples of both kinds of
particles on the velocity diagram and cannot be done for a single object where
the distribution of interlopers can be highly irregular. We therefore
stacked all 30 velocity diagrams into three composite ones in each projection
separately. All radii and velocities were rescaled by $r_{v}$ and $V_{v}$
respectively so that the mass dependence is factored out.
This procedure is commonly applied to the cluster data to improve
the statistics and to study the typical properties of
clusters which are expected to scale with mass (e.g. Mahdavi \& Geller 2004;
{\L}okas et al. 2006). Fig.~\ref{fig12} shows our three stacked velocity diagrams
in projection along the $x$, $y$ and $z$ axes.

For each of these diagrams, we calculate surface density profiles
and velocity histograms for all three kinds of interlopers (Fig.~\ref{fig13}).
It is clearly seen that unbound particles are the type
of interlopers with the most uniform surface density and velocity
distribution, whereas particles from the samples beyond $2r_{v}$ or
$r_{v}$ are considerably concentrated in the vicinity of the halo mean
velocity (see also Cen 1997, Diaferio et al. 1999 and Chen et al. 2006).
The uniformity of the distribution of interlopers on the velocity diagram
is the most natural assumption and has been used previously in the construction
of the probability distribution of interlopers  (Mahdavi \& Geller 2004).
We can see that this hypothesis agrees well with a distribution of unbound
particles.

\subsection{Estimation of the velocity dispersion profile with a uniform background
of interlopers}

\begin{figure}
\begin{center}
    \leavevmode
    \epsfxsize=8cm
    \epsfbox[50 50 400 300]{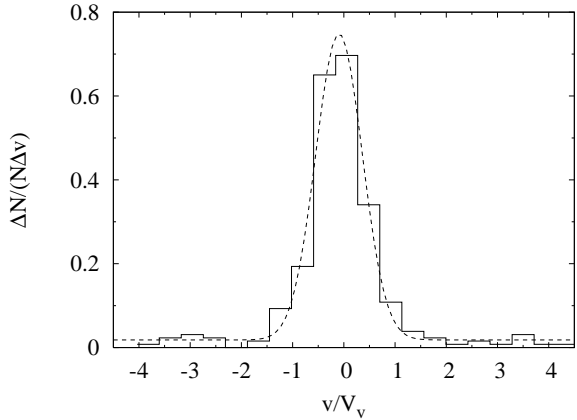}
\end{center}
\caption{The velocity histogram and the fitted probability distribution given by
(\ref{f_R}) for the outer bin of a stacked velocity diagram in projection along
the $x$ axis.}
\label{fig14}
\end{figure}

This method was originally introduced by Prada et al. (2003) in the
context of measuring the velocity dispersion profile of satellite galaxies.
Following these authors we assume that the distribution of particles is given by the
sum of a Gaussian part for halo particles and a constant background describing
interlopers. Let $f_{R}(v)\textrm{d}v$ be the conditional probability of finding
any type of particle in the infinitesimal range $[v,v+\textrm{d}v]$ at a given
radius $R$
\begin{equation}	\label{f_R}
	f_{R}(v)\textrm{d}v=\alpha(R)p_{v}\textrm{d}v + [1-\alpha(R)]
	\frac{\textrm{d}v}{2v_{\rm max}} ,
\end{equation}
where $p_{v}$ is the Gaussian distribution given by (\ref{p_v}) with
dispersion equal to $\sigma_{\rm los}(R)$ and the local mean velocity $\mu(R)$,
$v_{\rm max}$ is the maximum velocity available on the velocity diagram and
$\alpha(R)$ has a simple interpretation of the probability of finding a halo
particle at a given radius $R$. Note that in this case the
normalization condition is valid only for a fixed radius $R$. All
radius-dependent quantities are estimated in radial bins by fitting formula
(\ref{f_R}) to a velocity histogram in a given bin by minimizing the $\chi^{2}$
function. We use 10 radial bins with 300 data points in each of them.

\begin{figure}
\begin{center}
    \leavevmode
    \epsfxsize=8.5cm
    \epsfbox[50 50 480 500]{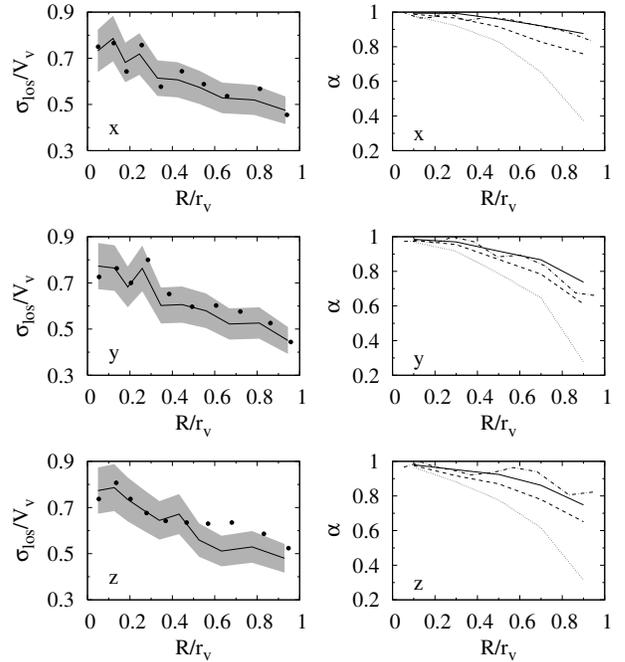}
\end{center}
\caption{Left column panels show dispersion profiles obtained by fitting
the probability distribution (\ref{f_R}) to velocity histograms for composite
haloes in different
radial bins (filled circles). Black lines plot dispersion profiles of
bound particles with shaded strips indicating the $3\sigma$ range of variability among
the velocity diagrams.
Right column panels show profiles of the probability $\alpha(R)$ measured
for particles beyond $r_{v}$ (dotted lines) or $2r_{v}$
(dashed lines) and for unbound particles (solid lines).
Dashed-dotted broken lines represent results of the fitting procedure.
Each row of panels corresponds to a different projection direction of the stacked
velocity diagram marked in the bottom left corner.}
\label{fig15}
\end{figure}

An example of the velocity histogram together with the fitted probability
distribution given by (\ref{f_R}) is shown in Fig.~\ref{fig14}. The dispersion
profiles obtained in the fitting procedure are shown in the left column of
Fig.~\ref{fig15} (filled circles). For comparison, we also plot velocity
dispersion profiles of bound particles with shaded areas
corresponding to $\pm 3\sigma$ departures from the mean. As we can see, all profiles
determined in the fitting procedure decline properly with radius and
trace well the dispersion profiles of bound particles.
Interestingly, some clearly overestimated values of $\sigma_{\rm los}$ and $\alpha$ probability
appear in the same range of radii for the case of projection along the $z$ axis which is an
effect of local irregularities in the velocity distribution of unbound particles.

It seems interesting to compare $\alpha(R)$ profiles
obtained in the fitting procedure with those directly measured from the
data for different types of interlopers.
Using frequency definition of probability $\alpha$ is expressed as
\begin{equation}	\label{alpha_R1}
	\alpha(R)=\frac{N_{g}(R)}{N_{g}(R)+N_{i}(R)} ,
\end{equation}
where $N_{g}$ and $N_{i}$ are numbers of halo particles and interlopers
defined by a given criterion respectively. In the right column panels of
Fig.~\ref{fig15} we plot $\alpha$ profiles estimated in radial bins with
interlopers defined as: unbound particles (solid lines), particles beyond
$2r_{v}$ (dashed lines) and $r_{v}$ (dotted lines). Profiles obtained in the
fitting procedure are indicated with broken dashed-dotted lines. We can clearly see
that, as expected, the fitted $\alpha$ profiles reproduce direct measurements
best for bound particles.

We therefore conclude that the favoured group of
interlopers in this approach consists of unbound particles, whereas
most particles from the outside of the virial region, but bound to a given
halo contribute to the Gaussian part of the distribution function
$f_{R}(v)$ and in fact are used in the estimation of the dispersion
profile. This situation is difficult to avoid; wishing to include
particles beyond $2r_{v}$ or $r_{v}$ as interlopers one would have
to introduce another Gaussian-like distribution of interlopers in velocity
space (see velocity histograms in Fig.~\ref{fig13}). This, however, would
cause strong degeneracy since both halo particles and interlopers would
be described by very similar distributions.

\subsection{Bayesian technique}

In this subsection we study the elegant approach of statistical treatment of
interlopers originally proposed by van der Marel et al. (2000) and
Mahdavi \& Geller (2004). This method is based on the Bayes technique which
allows us to determine the probability distribution in the parameter space of a
particular model given the measured data. Consider the parameter set ${\mathbi a}$
and data set $\{{\mathbi x}_{i}\}$.
Following Bayes theorem the probability of getting certain values of
parameters ${\mathbi a}$ given data sequence $\{{\mathbi x}_{i}\}$ is
\begin{equation}      \label{bayes}
	p({\mathbi a}|\{{\mathbi x}_{i}\})=\frac{p({\mathbi a})}{p(\{{\mathbi x}_{i}\})}
	\Pi_{i}p({\mathbi x}_{i}|{\mathbi a}) ,
\end{equation}
where $p({\mathbi a})$ is the prior on the parameters and
$p(\{{\mathbi x}_{i}\})$ takes care of normalization. The combination of
$p({\mathbi x}_{i}|{\mathbi a})$ on the right-hand side of equation (\ref{bayes})
is the likelihood, while $p({\mathbi a}|\{{\mathbi x}_{i}\})$ is the posterior
probability. Obviously, each of the probability distributions introduced above is
normalized to unity in the available part of the corresponding space
\begin{eqnarray}
	\int_{-\infty}^{+\infty}...\int_{-\infty}^{+\infty}
	p({\mathbi a}|\{{\mathbi x}_{i}\})\textrm{d}a_{1}...\textrm{d}a_{n} & =& 1\\
	\int_{-\infty}^{+\infty}...\int_{-\infty}^{+\infty}
	p({\mathbi x}_{i}|{\mathbi a})\textrm{d}x_{1}...\textrm{d}x_{n} & =& 1.
\end{eqnarray}

Van der Marel et al. (2000) proposed to restrict the considerations to the
velocity space (${\mathbi x}_{i}=v_{i}$). The probability $p(v_{i}|{\mathbi a})$ was
given by (\ref{f_R}) with a more detailed formula for $p_{v}$ and the assumption
that $\alpha(R)\approx\textrm{const}$, which was found to be roughly consistent
with the data. However, as we have seen in Fig.~\ref{fig15} the dependence of the
probability $\alpha $ on radius is in general not negligible. The most natural
way to take this fact into account is simply to consider the probability on the
whole projected space with ${\mathbi x}_{i}=({\mathbi R}_{i},v_{i})$
\begin{eqnarray}
	p(R_{i},v_{i}|({\mathbi a}_d,\alpha_{p}))\textrm{d}R\textrm{d}v
	& = & \alpha_{p} f(R_{i},v_{i}|{\mathbi a}_d)\textrm{d}R\textrm{d}v
	\label{p_Rv} \\
	& + & (1-\alpha_{p})\frac{R}{R_{\rm max}^{2}v_{\rm max}}
	\textrm{d}R\textrm{d}v ,  \nonumber
\end{eqnarray}
where $f(R,v)$ is the projected distribution function of halo particles, ${\mathbi a}_d$
is a set of dynamical parameters and $\alpha_{p}$ is an additional free
parameter describing the probability that a particle found at any radius and
with any velocity is a halo particle. The last term in (\ref{p_Rv}) describes
the uniform distribution of interlopers both in position on the sky and in
velocity space.

\begin{figure}
\begin{center}
    \leavevmode
    \epsfxsize=8.5cm
    \epsfbox[50 50 480 500]{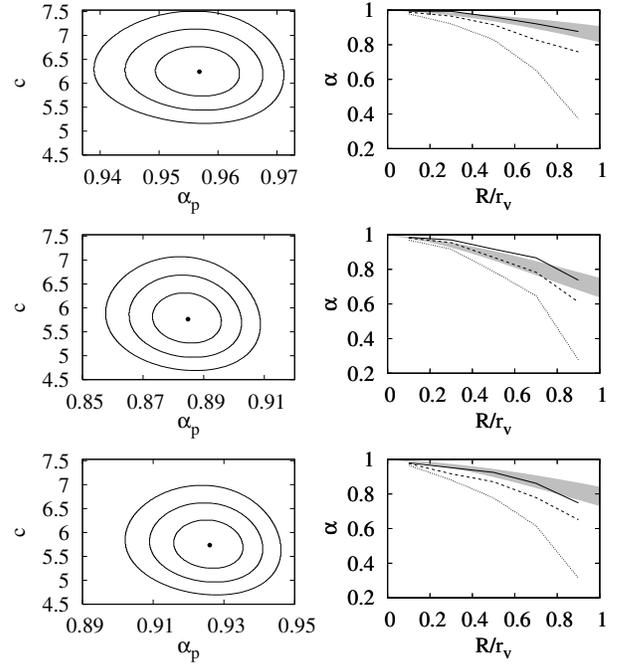}
\end{center}
\caption{Left column panels show the results of the Bayesian analysis in the form of
contours in the $c$-$\alpha_{p}$ parameter plane
corresponding to 68.3, 95.4 and 99.73 percent confidence levels. In the
right column we plot $\alpha(R)$ probability profiles calculated for different types of
interlopers: particles beyond $r_{v}$ (dotted lines) or $2r_{v}$ (dashed lines)
and unbound particles (solid lines).
Shaded areas indicate $\alpha(R)$ values corresponding to $c$ and
$\alpha_{p}$ parameters from the inside of the 99.73 percent probability contours.
The rows of panels from top to bottom are for the $x$, $y$ and $z$
projections respectively.}
\label{fig16}
\end{figure}

We performed the analysis of our three stacked haloes using the probability
distribution (\ref{p_Rv}) with $f(R_{i}, v_{i}|{\mathbi a})$ given for
simplicity by the formula (\ref{p_Mv}) parametrized by the $c$ parameter.
Since we are dealing with stacked haloes and we assume isotropic orbits
(as shown to be the case for early-type galaxies in Coma and other clusters,
see {\L}okas \& Mamon 2003; Biviano \& Katgert 2004), the only free parameters
in this analysis are the concentration and the probability $\alpha_{p}$.
Results of this analysis are shown in the left column of Fig.~\ref{fig16} in
the form of contours corresponding to 68.3, 95.4 and 99.73 percent confidence
levels drawn in the $c$-$\alpha_{p}$ parameter plane.

As we can see from Fig.~\ref{fig16}, the preferred values of $c$ lie slightly
below the value of the concentration parameter (estimated from the full 3D data)
for the composite cluster, $c=6.9$, which is probably due to our simplistic probability
distribution $f(R,v)$. However, we believe that the results concerning
$\alpha_{p}$ illustrate well the general features of interloper treatment in
this approach since the distribution of interlopers used in formula (\ref{p_Rv})
is independent of the dynamical model.

It is interesting to
compare $\alpha_{p}$ with abundances of different types of particles seen on
velocity diagrams and the mean value of $\alpha$ obtained in the previous
subsection. All these values are listed in
Table~\ref{stack}, where $F_{<v_{\rm esc}}$ and $F_{<2r_{v}}$ are the abundances of bound
particles and particles within $2r_{v}$ respectively. Comparing these results,
we notice an excellent agreement between $\alpha_{p}$ and $\bar{\alpha}$. This
fact is understandable since $\alpha$ was estimated in
radial bins with the same number of particles inside each of them. Consequently
one expects that the mean value of $\alpha$ measures the probability that a
particle randomly chosen from the velocity diagram is a halo particle which is exactly
the same probability as described by $\alpha_{p}$. We also find that
$\alpha_{p}$ is approximately equal to the abundance of bound particles. This
means that in this approach mainly unbound particles contribute
to the distribution of interlopers. Some departure from this rule can be
seen for the $z$-axis projection. However, if we take into
account that the velocity distribution of unbound particles in this case is
slightly peaked close to $v\sim V_{v}$ (see the bottom right panel of
Fig.~\ref{fig13}) the situation is clarified: some of these interlopers
are identified as halo particles increasing $\alpha_{p}$.

\begin{table}
\caption{Probabilities ($\alpha_{p}$ and $\bar{\alpha}$) of finding a halo
particle on the velocity diagram estimated via indirect methods of interloper
treatment. $F_{<v_{\rm esc}}$ and $F_{<2r_{v}}$ are fractions of bound particles and particles
within $2 r_v$ respectively.}
\label{stack}
\begin{center}
\begin{tabular}{ccccccc}
halo projection & $\alpha_{p}$        & $\bar{\alpha}$ & $F_{<v_{\rm esc}}$ & $F_{<2r_{v}}$\\
\hline
     &                                &                &            &      \\
$x$    & $0.957^{+0.014}_{-0.018}$      & 0.946  & 0.958      & 0.909\\
     &                                &                &            &      \\
$y$    & $0.885^{+0.022}_{-0.027}$      & 0.883  & 0.903      & 0.852\\
     &                                &                &            &      \\
$z$    & $0.926^{+0.020}_{-0.024}$      & 0.927   & 0.905      & 0.856\\
     &                                &                &            &      \\
\hline
\end{tabular}
\end{center}
\end{table}

Although we used here only the global probability $\alpha_{p}$ of finding a
halo particle in the velocity diagram, it is easy to reproduce the $\alpha(R)$
profile. In this case equation (\ref{alpha_R1}) has its counterpart in the
form
\begin{equation}	\label{alpha_R2}
	\alpha(R)=\frac{\Sigma_{M}(R,c)}{\Sigma_{M}(R,c)+\Sigma_{i}} ,
\end{equation}
where $\Sigma_{i}=\textrm{const}$ is the surface density of interlopers
related to $\alpha_{p}$ by
\begin{equation}
	\alpha_{p}=\frac{\int_{0}^{R_{\rm max}}2\pi R\Sigma_{M}(R,c)\textrm{d}R}
	{\int_{0}^{R_{\rm max}}2\pi R\Sigma_{M}(R,c)\textrm{d}R+\pi
	R_{\rm max}^{2}\Sigma_{i}}.
\end{equation}
Therefore given $\alpha_{p}$ and $c$ one is able to calculate $\alpha(R)$
profile. The results are plotted in the right column panels of Fig.~\ref{fig16}
in the form of shaded areas which indicate $\alpha(R)$ values calculated for
$c$ and $\alpha_{p}$ parameters from the inside of the probability contour
corresponding to 99.73 percent confidence level. We can clearly see that the area
available for $\alpha(R)$ includes a profile calculated for interlopers as
unbound particles.

We therefore confirm the results of the previous subsection
that the uniform distribution of interlopers in the velocity diagram is mainly
reproduced by unbound particles, whereas most of bound particles from the
outside of the virial sphere contribute to the distribution function of halo particles.
We emphasize
that only indirect methods are able to take into account the presence of low velocity interlopers
in the central part of the velocity diagram. Consequently, they can potentially
include all unbound particles in the interloper background, which is beyond
the reach of any direct scheme. For instance, applying $v_{\rm max}(1)$ method, the
most successful among direct ones,  to the composite velocity diagrams
we found that 21, 16 and 44 percent of unbound particles (for $x$, $y$ and $z$ projection
respectively) are not identified and remain in the final samples of halo members
resulting in the contamination on the level of 0.9, 1.7 and 4.4 percent respectively.


\section{Discussion}

We have studied different approaches to the treatment of interlopers in the analysis
of kinematic data for galaxy clusters. For the direct methods of interloper removal
their efficiency was measured by simple parameters: the fraction of removed interlopers $f_i$,
the fraction of removed members $f_g$ and the fraction of non-identified interlopers remaining
in the final samples $f_{h}$. The values of these parameters obtained by avergaing over 30
velocity diagrams studied here are given in Table~\ref{methods}. We can
see that the highest $f_i$ (73 percent for unbound particles)
is reached by $v_{\rm max}(1)$ method originally proposed by den Hartog
\& Katgert (1996) although all the other methods considered (except for $v_{\rm max}(2)$)
have $f_{i}$ on the level of
60 percent. The differences in mean $f_{i}$ between the methods are small compared
to the dispersion obtained by averaging the results over 30 velocity diagrams
which is about 20-30 percent in the case of unbound particles.
In all direct methods of interloper removal
$f_{i}$ equals approximately 40 percent and 20 percent respectively for particles beyond
$2r_{v}$ and $r_{v}$ so the methods are not efficient in removing them. This occurs because
particles from the close surroundings of the virial region are significantly concentrated around
the halo mean velocity and may in fact be good tracers of the halo potential.
All the methods show rather low fractions of members misidentified as interlopers
(below 2 percent for all types of interlopers with a dispersion of about 2 percent).
Furthermore, final samples of halo members include on average 2-4 percent of unbound
particles with dispersion of around 5 percent.

One may wonder if this is the most reliable way to compare the different methods. Since our
purpose is to maximize $f_i$ and minimize $f_g$ and $f_h$ one could construct some combination of these
parameters assigning them different weights; we have not done this in order not to obscure the
picture. One can also ask whether removing 60 percent of
interlopers by one method cannot be better than removing 70 percent by another when it comes
to the final estimation of the cluster parameters. It could happen that a smaller number of
removed interlopers could lead to better estimates of mass because the interlopers were actually those
causing the largest bias while larger number of interlopers can in principle lead to more biased
results because the less significant interlopers were removed. Fortunately, this is not the case,
the interlopers removed by the different methods are mostly the same, the difference is usually in
the border-line particles which do not significantly contribute to the bias. We have also verified
that the final, cleaned samples depend very weakly on the initial cut-off in velocity: reducing
the cut-off to 3000 km s$^{-1}$ from 4000 km s$^{-1}$ with respect to the cluster mean produces
almost identical final samples differing by only 1-2 particles in a few out of 30 velocity diagrams.

Apart from the quantitative measures presented in Table~\ref{methods} the choice between different
methods of interloper removal is largely a matter of subjective preference. One could try to judge
the methods by how strong their underlying assumptions are but the methods rely on
such different assumptions that it is difficult to compare them. Another possibility is to look at
their convergence. Here again $v_{\rm max}(1)$ method is recommended: it
always converges, contrary e.g. to second most effective method $M_{P}/M_{VT}$ based on the ratio of mass
estimators where the procedure has to be stopped at a rather arbitrary point. Furthermore, as shown
by van Haarlem et al. (1997), $v_{\rm max}(1)$ approach leads to the data sample which reproduces real
velocity dispersion better than the one obtained after applying $3\sigma_{\rm los}$ algorithm in the
original form proposed by Yahil \& Vidal (1977). It is also worth noting
that $v_{\rm max}(1)$ method does not involve any parameters or characteristic scales which could
restrict its application to cluster-size objects. It has been recently successfully applied to
tidally stripped
dwarf spheroidal galaxies where it allows to clean the stellar kinematic samples from interlopers
originating from the tidal tails and the Milky Way (Klimentowski et al. 2006).

An ultimate verification of the methods would of course come from the dynamical modelling
performed on the cleaned samples. Here, however,
other issues come into play: the quality of the final results of the modelling depends also on the
actual method used and particular properties of an object (whether it is in equilibrium, whether it
is spherically symmetric, how well is the kinematics sampled). These sources of error may have
more impact on the final result than the {\em differences} between the methods of interloper removal
studied here (see the recent study by Biviano et al. (2006) who have analyzed, after interloper removal
with $v_{\rm max}(1)$ method, different observational effects affecting the determination of cluster mass and
velocity dispersion). However, if interlopers were not removed at all the result would be very strongly
biased and the contamination could become the main source of error. In any case it would be
very difficult to disentangle the effect of interlopers from other sources of uncertainties.

Comparison between many modelling approaches possible is beyond the scope of this paper
and it was not our purpose to provide here such a final answer. For the study of the performance
of one of the dynamical analysis methods based on fitting the velocity dispersion and kurtosis
profiles for clusters we refer the reader to Sanchis et al. (2004) and {\L}okas
et al. (2006). Examples of such full dynamical modelling, including the interloper removal as an
important first step, in the case of Abell 576 and other clusters with significant background
contamination can be found in Wojtak \& {\L}okas (2006).

For the purpose of the present study we performed a simple
test taking the cleaned particle samples for our 30 velocity diagrams obtained with $v_{\rm max}(1)$ method
and fitting the velocity dispersion profiles to the solutions of the Jeans equation assuming isotropic
orbits and estimating the virial masses and concentrations. Averaging over 30 diagrams we find that
the ratio of the estimated virial mass to the real one measured from the 3D information is
$M_v/M_{v, {\rm true}} = 0.86 \pm 0.23$ while for concentrations we get $c/c_{\rm true} = 1.76 \pm 1.01$.
Interestingly, for the samples of bound particles the results are very similar:
$M_v/M_{v, {\rm true}} = 0.85 \pm 0.18$, $c/c_{\rm true} = 1.91 \pm 1.31$. These values should be
compared to the parameters obtained for the samples of all particles (without application of any
interloper removal scheme): $M_v/M_{v, {\rm true}} = 1.64 \pm 0.92$, $c/c_{\rm true} = 0.34 \pm 0.31$.
We therefore conclude that
using the samples cleaned with $v_{\rm max}(1)$ method is equivalent to working with only
bound particles. The significant bias still present in both cases, especially for concentration, can
be explained by departure from isotropy of particle orbits in the simulated haloes which should be
taken into account in more complete modelling.

We have also considered indirect methods of interloper treatment where their presence is accounted
for in a statistical way. Here we have found that, contrary to the assumptions of van der Marel
et al. (2001), the probability of a given galaxy being an interloper is not independent of the projected
radius but increases with it. We have also verified the applicability of the approach of fitting a
Gaussian plus a constant to the velocity distribution in galaxy clusters as a method to account for
interlopers in estimating the velocity dispersion profile originally proposed by Prada et al. (2003)
to study the velocity distribution of satellites around giant galaxies.

The main disadvantage of the
indirect methods is that in order to reliably estimate the parameters they require large kinematic
samples which can only be obtained by stacking data coming from many objects. This procedure, although
commonly used, can be dangerous, because it is not clear how the distances and velocities should be
scaled. While in the case of distances the choice of the virial radius as the scaling parameter
seems rather obvious, in the
case of velocities it is less clear whether one should use velocity dispersion, circular velocity
at the virial radius or perhaps the maximum circular velocity and still all would be subject to
uncertainties due to modelling of single clusters. In addition, the underlying assumption of indirect
methods is that the background of interlopers is uniform which is never exactly
the case due to clustering.


\section*{Acknowledgements}

We wish to thank our referee, A. Biviano, for constructive comments which helped to improve
the paper. Computer simulations used in this work were performed at the
Leibnizrechenzentrum (LRZ) in Munich.
RW and E{\L} are grateful for the hospitality of Astrophysikalisches
Institut Potsdam, Institut d'Astrophysique de Paris and
Instituto de Astrof{\'\i}sica de Andalucia where part of this work was done.
RW acknowledges the summer student program at Copernicus Center.
This work was partially supported by the Polish Ministry of Scientific
Research and Information Technology under grant 1P03D02726 as well as the
Jumelage program Astronomie France Pologne of CNRS/PAN, the Polish-Spanish
exchange program of CSIC/PAN and the Polish-German exchange program of
Deutsche Forschungsgemeinschaft (DFG).


\end{document}